\documentclass[10pt,prl,preprintnumbers,floatfix,aps,nofootinbib,showpacs,twocolumn,superscriptaddress,noshowpacs,preprintnumbers]{revtex4} 

\usepackage{amsmath}
\usepackage{graphicx}
\usepackage[dvipsnames]{xcolor}
\usepackage{dcolumn}
\usepackage{bm}

\usepackage{url}
\usepackage{hyperref}
\usepackage{soul} 
\usepackage[normalem]{ulem} 
\usepackage{bm}
\usepackage{wasysym}
\usepackage{verbatim}
\usepackage{amsfonts}
\usepackage{ragged2e}

\definecolor{DARKBLUE}{HTML}{00008b}
\hypersetup{colorlinks=true,
urlcolor=DARKBLUE
,anchorcolor=DARKBLUE
,citecolor=DARKBLUE
,filecolor=DARKBLUE
,linkcolor=DARKBLUE
,menucolor=DARKBLUE
,linktocpage=true,
pdfproducer=medialab
}

\usepackage{array} 

\begin{document}
\preprint{IFT-UAM/CSIC-24-169}

\title{Outliers in DESI BAO:  robustness and cosmological implications}

\author{Domenico Sapone}%
\email{domenico.sapone@uchile.cl}

\affiliation{%
Departamento de Física, FCFM, Universidad de Chile, Santiago, Chile.}%

\author{Savvas Nesseris}%
\email{savvas.nesseris@csic.es}

\affiliation{Instituto de Física Teórica UAM-CSIC, Universidad Autónoma de Madrid, Cantoblanco, 28049 Madrid, Spain.}

\date{\today}

\begin{abstract}
We apply an Internal Robustness (iR) analysis to the recently released Dark Energy Spectroscopic Instrument (DESI) baryon acoustic oscillations  dataset. This approach examines combinations of data subsets through a fully Bayesian model comparison, aiming to identify potential outliers, subsets possibly influenced by systematic errors, or hints of new physics. Using this approach, we identify three data points at $z= 0.295,\,0.51,\,0.71$ as potential outliers. Excluding these points improves the internal robustness of the dataset by minimizing statistical anomalies and enables the recovery of $\Lambda$CDM predictions with a best-fit value of  $w_0 = -1.050 \pm 0.128$ and $w_a = 0.208 \pm 0.546$. These results raise the intriguing question of whether the identified outliers signal the presence of systematics or point towards new physics.

\end{abstract}

\maketitle

\textbf{\emph{Introduction.}} 
Despite significant progress, our understanding of the Universe remains incomplete, and one of the most surprising discoveries of the last thirty years is that its expansion is accelerating. First observed in 1998 through the study of Type Ia supernovae \cite{SupernovaSearchTeam:1998fmf, SupernovaCosmologyProject:1998vns}, this phenomenon has since been confirmed by numerous other methods and datasets \cite{SDSS:2003eyi, SDSS:2005xqv, SDSS:2006lmn, Stern:2009ep, WMAP:2012nax, BOSS:2012dmf, SDSS:2014iwm, BOSS:2014hhw, Moresco:2016mzx, BOSS:2016wmc, eBOSS:2020mzp, DES:2017qwj, Planck:2018vyg, ACT:2020gnv, DES:2021wwk, Brout:2022vxf, DES:2024jxu}.  In classical physics, any form of matter or energy would naturally slow the expansion of the Universe due to gravitational attraction. Acceleration, however, introduces the need for a new explanation, leading to the concept of Dark Energy (DE), a mysterious component that exerts a repulsive force on cosmological scales. Within the $\Lambda$CDM framework, DE is explained as a cosmological constant $\Lambda$, which enters as a constant term into Einstein’s equations. This elegant solution implies that the total energy budget of the Universe is dominated by a form of DE. 

Despite its success, this model raises significant theoretical challenges, such as the fine-tuning and coincidence problems, see Refs.  \cite{Sapone:2010iz,Martin:2012bt, Perivolaropoulos:2021jda}. As a result, the nature of DE remains one of the greatest challenges in modern physics. It is no surprise that understanding cosmic acceleration remains a top priority for ongoing \cite{DESI:2024mwx, EUCLID:2011zbd, Euclid:2021icp, Euclid:2024yrr} and forthcoming large scale structure (LSS) galaxy surveys \cite{LSSTDarkEnergyScience:2018jkl} and early time physics \cite{SimonsObservatory:2018koc}.

While previous surveys have consistently supported $\Lambda$CDM as the preferred cosmological model, recent results from DESI introduced intriguing hints suggesting a potential strong variation in the equation of state for DE \cite{DESI:2024mwx}. These results generated significant attention from the community, reflecting the excitement surrounding these findings.

To explore potential deviations from the cosmological constant, the DESI analysis adopted the $w_0w_a$CDM model, an extension of $\Lambda$CDM that allows the DE equation of state to vary with time. First proposed by Chevallier, Polarski, and Linder \cite{Chevallier:2000qy, Linder:2002et}, this parameterization expresses the equation of state as $w(a) = w_0 + w_a(1-a)$, where $w_0$ represents its current value, and $w_a$ characterizes its rate of change. Unlike the constant $w = -1$ assumed in $\Lambda$CDM, the $w_0w_a$CDM model offers a more flexible framework to probe the nature of DE.

The question now is: do the DESI results hint at the emergence of a new cosmological paradigm, challenging the well-established  $\Lambda$CDM model?

Some recent studies have  examined this claim. For instance, \cite{Colgain:2024xqj} analyzed the DESI dataset and identified a $\sim 2\sigma$ discrepancy with Planck-$\Lambda$CDM predictions. Their work, focusing on the Luminous Red Galaxy (LRG) sample at an effective redshift of $0.51$, revealed a higher matter density parameter ($\Omega_{\rm m,0} \sim 0.668$) across all redshift bins, influencing the preference for a DE equation of state parameter $w_0 > -1$ in the $w_0w_a$CDM model. Furthermore, in \cite{Chudaykin:2024gol}, the authors tested the consistency of the DESI results by replacing the DESI distance measurements at $z < 0.8$ with the SDSS BAO measurements over the same redshift range \cite{eBOSS:2020mzp}. Assuming a $w_0 w_a$CDM model, they found the equation of state parameters to be consistent with $\Lambda$CDM. 

Guided by these works, we explore the internal robustness of DESI BAO measurements, fully aware that excluding data points is a sensitive decision that may raise concerns about the loss of valuable information or potential misinterpretation of the dataset's reliability. In this letter, we apply the robust statistical analysis proposed by Amendola et al. \cite{Amendola:2012wc} and we find that three data points in the DESI dataset can be considered outliers, with two being the most problematic. In addition, excluding these two points allowed us to recover the $\Lambda$CDM model as the best-fit cosmology, with parameters $w_0 = -0.956 \pm 0.125$ and $w_a = -0.166 \pm 0.561$.

\textbf{\emph{Theory.}} Here we present the basic theoretical framework of our analysis. The Hubble parameter in a flat $w_0 w_a$CDM universe (ignoring neutrinos and radiation at late times), is given by
\begin{equation}
	\frac{H(a)^2}{H_0^2} = \Omega_{m,0}(1+z)^{3}+(1-\Omega_{m,0})(1+z)^{3(1+\hat{w}(z))} \,,
    \label{eq:hubble-parameter}
\end{equation}
where $H_0$ is the Hubble constant, and $\Omega_{m,0}$ is the present day value of the matter density parameter and $z$ is the redshift. The equation of state parameter of DE is
\begin{equation}
    \hat{w}(z) =  \frac{1}{\log(1+z)}\int_{0}^{z}\frac{w(z')}{1+z'}\,{\rm d}z'\, ,
\end{equation}
where we use the CPL parameterization.

The transverse comoving distance is defined as:
\begin{equation}
D_{\rm M}(z)=c\int_{0}^{z}{\frac{{\rm d}z'}{H(z')}}\,,
    \label{eq:comoving-distance}
\end{equation}
where $c$ is the speed of light. The Hubble distance and the volume distances are defined respectively as: 
\begin{eqnarray}
    D_{H}(z) &=& \frac{c}{H(z)}\,,\\
    D_{\rm V}(z) &=&  \Big[z\,D_{\rm M}^2(z) \, D_{H}(z)\Big]^{1/3}\,.
    \label{eq:distances}
\end{eqnarray}
Finally, for standard early-time physics, the drag-epoch sound horizon can be written as \cite{DESI:2024mwx, Brieden:2022heh}
\begin{equation}
    r_{\rm d} = 147.05 \left(\frac{\omega_{\rm m}}{0.1432}\right)^{-0.23}\,\left(\frac{N_{\rm eff}}{3.04}\right)^{-0.1}\, \left(\frac{\omega_{\rm b}}{0.02236}\right)^{-0.13},\nonumber
\end{equation}
where $\omega_{\rm m} = \Omega_{\rm m,0}h^2 = 0.14277$, $\omega_{\rm b} = \Omega_{\rm b,0}h^2 = 0.02233$, $N_{\rm eff} = 3.044$, and $h=0.67$ with these values adopted from Planck \cite{Aghanim:2018eyx}.

The BAO signal arises from the physics of sound waves in the photon-baryon plasma propagating in the early Universe, which imprint a characteristic scale ($r_{\rm d}$) in the matter distribution. The BAOs act like a cosmic ruler to measure distances in the Universe. By observing how galaxies are separated, we can measure distances across (perpendicular to our line of sight $D_{\rm M}$) and along (parallel to our line of sight $D_{H}$) the Universe. The transverse distance tells us about the Universe’s size at a given redshift, while the radial distance gives information about its expansion rate. When data quality is lower, an average of these distances is used ($D_{\rm V}$).

\textbf{\emph{Internal Robustness.}} The key equation used to identify outliers within a dataset is derived from the concept of internal robustness, first proposed in \cite{Amendola:2012wc} and later tested in \cite{Sagredo:2018ahx} for growth of matter density data. In the latter, the authors demonstrated that this approach is both stable and highly sensitive to dataset selection. 

The Bayesian evidence is the central concept in the internal robustness analysis and it is mathematically expressed as:
\begin{equation}
\mathcal{E}({\bf x}|M) = \int L({\bf x}|\bm{\theta}^{M})\pi(\bm{\theta}^M)\,\mathrm{d}\bm{\theta}^M\,,
\end{equation}
where ${\bf x} = (x_1, x_2, \dots, x_N)$ represents the set of $N$ observational data points, and $\bm{\theta}^M = (\theta_1, \theta_2, \dots, \theta_n)$ are the $n$  parameters for the model $M$. Here, $L({\bf x}|\bm{\theta}^{M})$ denotes the likelihood function, and $\pi(\bm{\theta}^M)$ is the prior probability distribution over the model parameters. Using Bayes' theorem, the posterior probability of model $M$ given the data can be formulated as:
\begin{equation}
\mathcal{P}(M|{\bf x}) = \frac{\mathcal{E}({\bf x}|M)\pi(M)}{\pi({\bf x})}\,.
\end{equation}
To compare the probabilities of two competing models, we calculate their ratio:
\begin{equation}
\frac{\mathcal{P}(M_1|{\bf x})}{\mathcal{P}(M_2|{\bf x})} = \mathcal{B}_{12} \frac{\pi(M_1)}{\pi(M_2)}\,,
\end{equation}
where the Bayes factor $\mathcal{B}_{12}$ is defined as:
\begin{equation}
\mathcal{B}_{12} = \frac{\mathcal{E}({\bf x}|M_1)}{\mathcal{E}({\bf x}|M_2)}\,.
\end{equation}
If we assume equal prior probabilities for the two models, the Bayes factor alone determines which model is favored by the data: $\mathcal{B}_{12} > 1$ ($<1$) supports $M_1$ ($M_2$).

A robustness test introduces an additional assumption: the data is assumed to originate from two distinct distributions. This is crucial for two reasons. First, it allows the total evidence to be expressed as the product of individual evidences, and second, it provides a means to assess data reliability. As a consequence, we partition the dataset into subsets $\{{\bf x_1}, {\bf x_2}\}$ associated with models $M_1$ and $M_2$. The Bayes factor is, then, given by:
\begin{equation}
\mathcal{B}_{12} = \frac{\mathcal{E}({\bf x}|M_1)}{\mathcal{E}({\bf x_1}|M_1)\mathcal{E}({\bf x_2}|M_2)}\,.
\end{equation}
This leads to the definition of internal robustness:
\begin{equation} \text{iR}_{12} = \log\mathcal{B}_{12} = \log\left(\frac{\mathcal{E}({\bf x}|M_1)}{\mathcal{E}({\bf x_1}|M_1)\mathcal{E}({\bf x_2}|M_2)}\right)\,. 
\label{eq:internal-robustness} 
\end{equation} 
This formalism enables the evaluation of whether subsets of the data align with a given cosmological model or are influenced by systematic effects within the survey. Internal Robustness (iR) systematically compares the Bayesian evidence of subsets to assess dataset consistency. By dividing the data into smaller groups and calculating the evidence for each, iR is able to identify deviations that may indicate potential outliers—data points that disproportionately impact overall results. 

The assumption of distinct models in Eq.~\eqref{eq:internal-robustness} is not strictly required and, in this work, we apply the same cosmological model to both subsets. Specifically, we adopt the $w_0 w_a$CDM framework and parameterize it as $\bm{\theta} = (w_{0}, w_{a})$. Henceforth, we omit the superscript $M$.

Finally, we adopt uniform flat priors for both parameters. Specifically, $w_0$ is assigned a prior range of $[-4, -1/3]$, while $w_a$ is constrained to $[-4, 4]$. The upper limit of the prior on $w_0$ is derived from second Friedmann equation, which indicates that, for the Universe to undergo to a phase of accelerated expansion, the pressure must satisfy $p < -\rho/3$. This translates to a condition on the equation of state parameter at the present epoch: $w(a=1) = w_0 < -1/3$ if only DE acts at late times\footnote{We remind that the Einstein equations rely on the total energy density, which turns into the condition $p_{\rm tot}/\rho_{tot} = p_{\rm DE}/(\rho_{\rm DE}+\rho_{\rm m}) = w(a) \Omega_{\rm DE}(a)<-1/3$.}. The choice of the prior range for $w_a$, however, is less physically motivated since there is no definitive theoretical framework to predict the evolution of DE. As a result, we choose a  broad range to allow possible variations.

\begin{table}
\centering
\begin{tabular}{|c|c|c|c|c|}
\hline 
${\bf P}$ & \textbf{Redshift} &  \textbf{Measurements} & \textbf{Type} & \textbf{Tracer} \\    
\hline 
\hline 
1 & 0.295 & 7.93 & $D_{\rm V}/r_{\rm d}$ & BGS \\
\hline 
2 & 0.51 & 13.62 & $D_{\rm M}/r_{\rm d}$ & LRG1 \\
\hline 
3 & 0.51 & 20.98 & $D_{\rm H}/r_{\rm d}$ & LRG1 \\
\hline 
4 & 0.71 & 16.85 & $D_{\rm M}/r_{\rm d}$ & LRG2 \\
\hline 
5 & 0.71 & 20.08 & $D_{\rm H}/r_{\rm d}$ & LRG2 \\
\hline 
6 & 0.93 & 21.71 & $D_{\rm M}/r_{\rm d}$ & LRG3+ELG1 \\
\hline 
7 & 0.93 & 17.88 & $D_{\rm H}/r_{\rm d}$ & LRG3+ELG1 \\
\hline 
8 & 1.32 & 27.79 & $D_{\rm M}/r_{\rm d}$ & ELG2 \\
\hline 
9 & 1.32 & 13.82 & $D_{\rm H}/r_{\rm d}$ & ELG2 \\
\hline 
10 & 1.49 & 26.07 & $D_{\rm V}/r_{\rm d}$ & QSO \\
\hline 
11 & 2.33 & 39.71 & $D_{\rm M}/r_{\rm d}$ & Lya QSO \\
\hline 
12 & 2.33 & 8.52 & $D_{\rm H}/r_{\rm d}$ & Lya QSO \\
\hline 
\end{tabular}
\caption{DESI Year-1 data used in the analysis, from \cite{DESI:2024mwx}. The table includes the data point index ${\bf P}$, corresponding redshift, measurement, the type of measurement ($D_{\rm V}/r_{\rm d}$, $D_{\rm M}/r_{\rm d}$, or $D_{\rm H}/r_{\rm d}$), and the tracer type (e.g., BGS, LRG, ELG, QSO, Lya QSO).}
\label{tab:desi_y1_data}
\end{table}

\textbf{\emph{Data.}} The data used in this analysis are derived from the DESI Year-1 paper \cite{DESI:2024mwx} and are summarized in Table~\ref{tab:desi_y1_data}.  The dataset consists of 12 data points, with five of them correlated. These correlated measurements are at redshifts $z=0.51, 0.71, 0.93, 1.32, 2.33$. The corresponding covariance matrix, which quantifies these correlations, is reported in Table~\ref{tab:covariance_matrices}.

\begin{table}[h!]
\setlength{\extrarowheight}{10pt} 
\centering
\begin{tabular}{|c|c|} 
\hline
\textbf{Redshift } & \textbf{Covariance Matrix $C(z)$} \\ 
\hline
\hline
0.295 &  $ 0.0225 $ \\ \hline
0.51 & 
\(\begin{pmatrix}
0.0625 & -0.0678625 \\
-0.0678625 & 0.3721
\end{pmatrix}\) \\ \hline
0.71 & 
\(\begin{pmatrix}
0.1024 & -0.08064 \\
-0.08064 & 0.36
\end{pmatrix}\) \\ \hline
0.93 & 
\(\begin{pmatrix}
0.0784 & -0.038122 \\
-0.038122 & 0.1225
\end{pmatrix}\) \\ \hline
1.32 & 
\(\begin{pmatrix}
0.4761 & -0.12867 \\
-0.12867 & 0.1764
\end{pmatrix}\) \\ \hline
1.49 & $ 0.4489  $ \\ \hline
2.33 & 
\(\begin{pmatrix}
0.8836 & -0.07622 \\
-0.07622 & 0.0289
\end{pmatrix}\) \\ \hline
\end{tabular}
\caption{The covariance matrices 
$C(z)$ corresponding to the DESI BAO measurements at various redshifts.}
\label{tab:covariance_matrices}
\end{table}

Since our analysis relies on the evidence we need to define our likelihood. We build it as the classical $\chi^2$ as 
\begin{equation}
    L({\bf x}|\bm{\theta}^{M})\propto e^{-\chi^2} = e^{-\frac12 \mathcal{D}^TC^{-1}\mathcal{D}},\nonumber
\end{equation}
where $\mathcal{D} = D-M$ (measurement - model) is the data vector and $C$ the covariance matrix.  

To systematically partition the dataset, we divide them into two subsets. Initially, the first subset contains a single data point, while the second subset includes the remaining 11 points. All possible combinations of these partitions are generated, ensuring that each data point appears in the first subset exactly once. The size of the first subset is then  increased by one at each step, reducing the size of the second subset accordingly. This iterative process continues until both subsets contain an equal number of points (6 each), due to the symmetry in the partition in Eq.~\eqref{eq:internal-robustness}.

In our analysis, there are $2^{N-1} - 1 = 2047$ possible combinations, and we analyze all subsets. Since the total number of cases is manageable and the likelihood involves only two parameters ($w_0-w_a$), we chose to calculate the full grid and evaluate the evidences for all combinations. We divide the grid into $350$ steps for each parameter, resulting in a total of $122,500$ likelihood evaluations. Once the likelihood is computed, we perform a double integration over the parameters $w_0$ and $w_a$ to compute the evidence. The result is then normalized by the prior volume $V_{\rm p}$, to ensure the evidence is properly scaled. To ensure reliability, we cross-checked our results using alternative methods, including independent Bayesian evidence codes, such as \texttt{MCEvidence} \cite{Heavens:2017afc}, and grid sampling techniques. These tests confirmed the accuracy of our code, ensuring that our conclusions are not dependent on specific computational approaches.

\begin{figure}
\centering 
\includegraphics[scale=0.34]{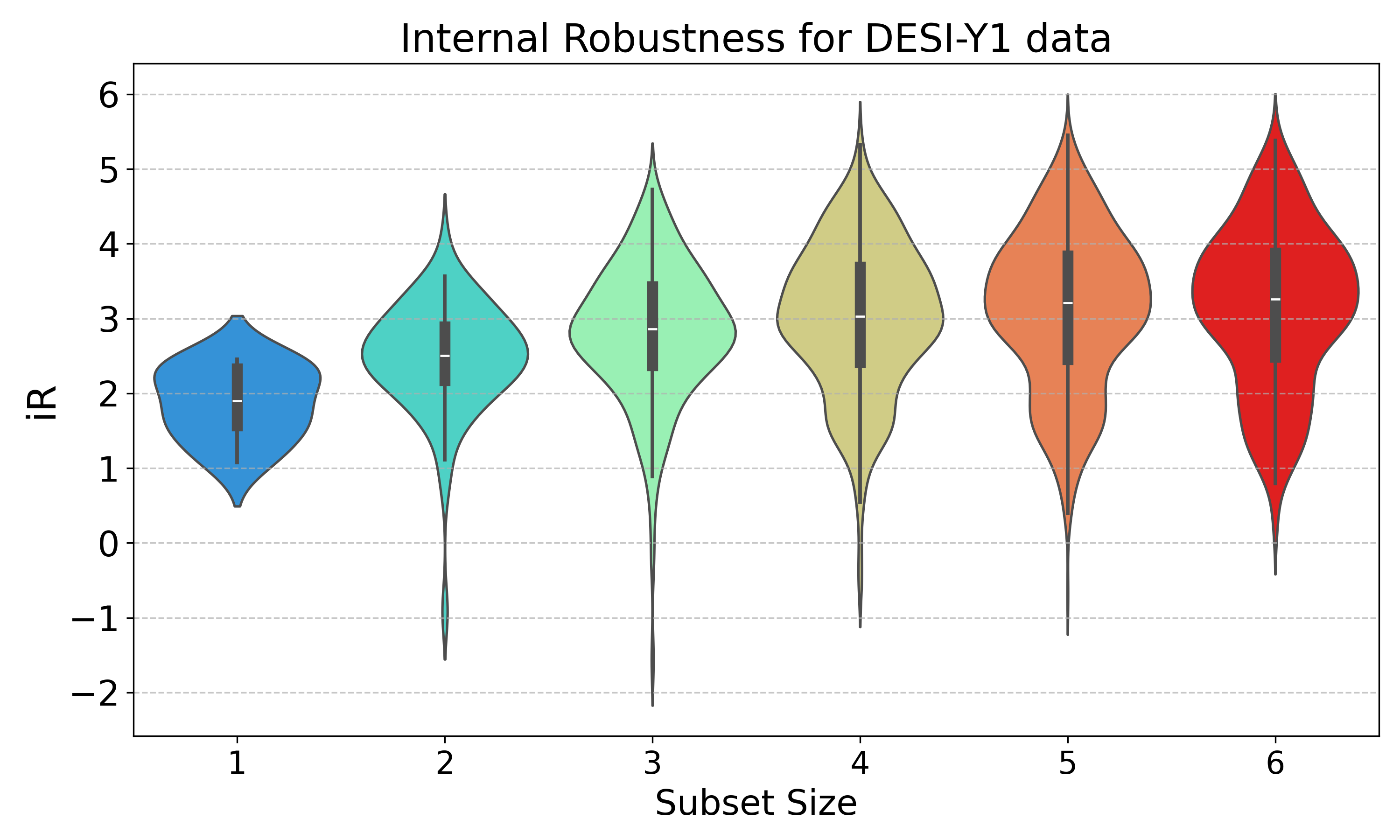}
\caption{\justifying Violin plots illustrating the internal robustness (iR) parameter as a function of subset size. Wider sections represent higher data concentration, while narrower sections highlight less frequent values. The long tails fie $\textbf{SSS} = 2$ and $\textbf{SSS} = 3$ indicate potential outliers in the dataset.}
\label{fig:ir}
\end{figure}

\textbf{\emph{Results.}} In this analysis, we fix all cosmological parameters,  i.e. the matter and baryon density parameters $\Omega_\mathrm{m,0}$ and $\Omega_\mathrm{b,0}$, except for $w_0$ and $w_a$,  which are allowed to vary freely. This choice is deliberate, as our primary goal is to verify the internal consistency and robustness of the BAO measurements derived from DESI data. Fixing the baryon density $\Omega_\mathrm{b,0}$ is equivalent to using the BBN prior, while fixing the matter density $\Omega_\mathrm{m,0}$ is equivalent to assuming the CMB prior on this parameter. The reason for this choice is that the DESI data alone cannot constrain all cosmological parameters at the same time, hence, focusing solely on $w_0$ and $w_a$ provides a controlled and clean framework to evaluate the reliability of the BAO data regarding the possible variations of the DE equation of state parameter $w(z)$. 

To visually report our results, we use the violin plot, which is a visualization tool that combines features of a box plot and a kernel density plot to represent the distribution of data across one or more categories. Wider sections indicate higher data concentrations, while narrower sections show rarer values. Inside the violin, the box plot highlights key summary statistics: the median (denoted by the horizontal white line), the interquartile range (represented by the vertical thick black line), and whiskers (represented by the vertical thin black line) that capture the spread of the data, excluding outliers. Additionally, the violin plot may include tails extending beyond the bulk of the data, which represent rare occurrences or potential outliers.
In Fig.~\ref{fig:ir}, we present our main result of the iR analysis using DESI data. 

As evident from Fig.~\ref{fig:ir}, the subsets of size $\textbf{SSS} = 2$ and $\textbf{SSS} = 3$ are most significantly affected, indicating that certain combinations of the data points exhibit statistical anomalies. In particular: 
\begin{itemize}
\item $\textbf{SSS} = 2 $: The distribution begins to widen, and the presence of outliers suggests that removing two specific data points has a noticeable impact on robustness. This could point to the existence of potential outlier data points within the dataset.
\item $\textbf{SSS} = 3$: The distribution becomes even wider, with more prominent outliers. This indicates that as more data points are excluded, the likelihood of encountering subsets with a larger effect on robustness increases. The presence of negative outliers for subsets of size $\textbf{SSS} = 3$, and to a lesser extent $\textbf{SSS} = 4$, reveals combinations that lead to significantly reduced robustness.
\end{itemize}
For higher $\textbf{SSS}$ values, the impact decreases, as the influence of the four (or more) particularly non-robust data points is mitigated when mixed with the larger dataset. Their individual effects are effectively diluted by the overall consistency of the remaining data. This is particularly evident for $\textbf{SSS} = 6$, where the distribution appears broad but no significant deviations are observed.  For $\textbf{SSS}=1$, the distribution remains narrow and well-behaved, indicating that removing a single data point does not significantly affect the results: a single data point rarely dominates the overall evidence unless it is an extreme outlier. 

Specifically, we found that for the case of ${\bf SSS}= 2$, the potential outliers are the third and fourth data points (denoted as ${\bf P}= 3$ and $4$). Similarly, for the case of ${\bf SSS}= 3$, the possible outliers are ${\bf P}= 1$, $3$, and $4$.

\begin{table}[h!]
\centering
\begin{tabular}{|c c c c c|}
\hline
Case & $\quad $ & $w_0$  & $\quad $ & $w_a$   \\ 
\hline
\hline
Full DESI &  & $-0.764 \pm 0.109$ & & $-0.874 \pm 0.539$  \\ 
\hline
SSS = 2  & & $-0.956 \pm 0.125$ & & $-0.166 \pm 0.561$  \\ \hline
SSS = 3  & & $-1.050 \pm 0.128$ & & $0.208 \pm 0.546$  \\ \hline
\end{tabular}
\caption{Best-fit parameters and $1\sigma$ errors for the full DESI dataset and for ${\bf SSS} =2$ and ${\bf SSS} =3$.}
\label{tab:best_fit_parameters}
\end{table}

\begin{figure}
\centering 
\includegraphics[scale=0.43]{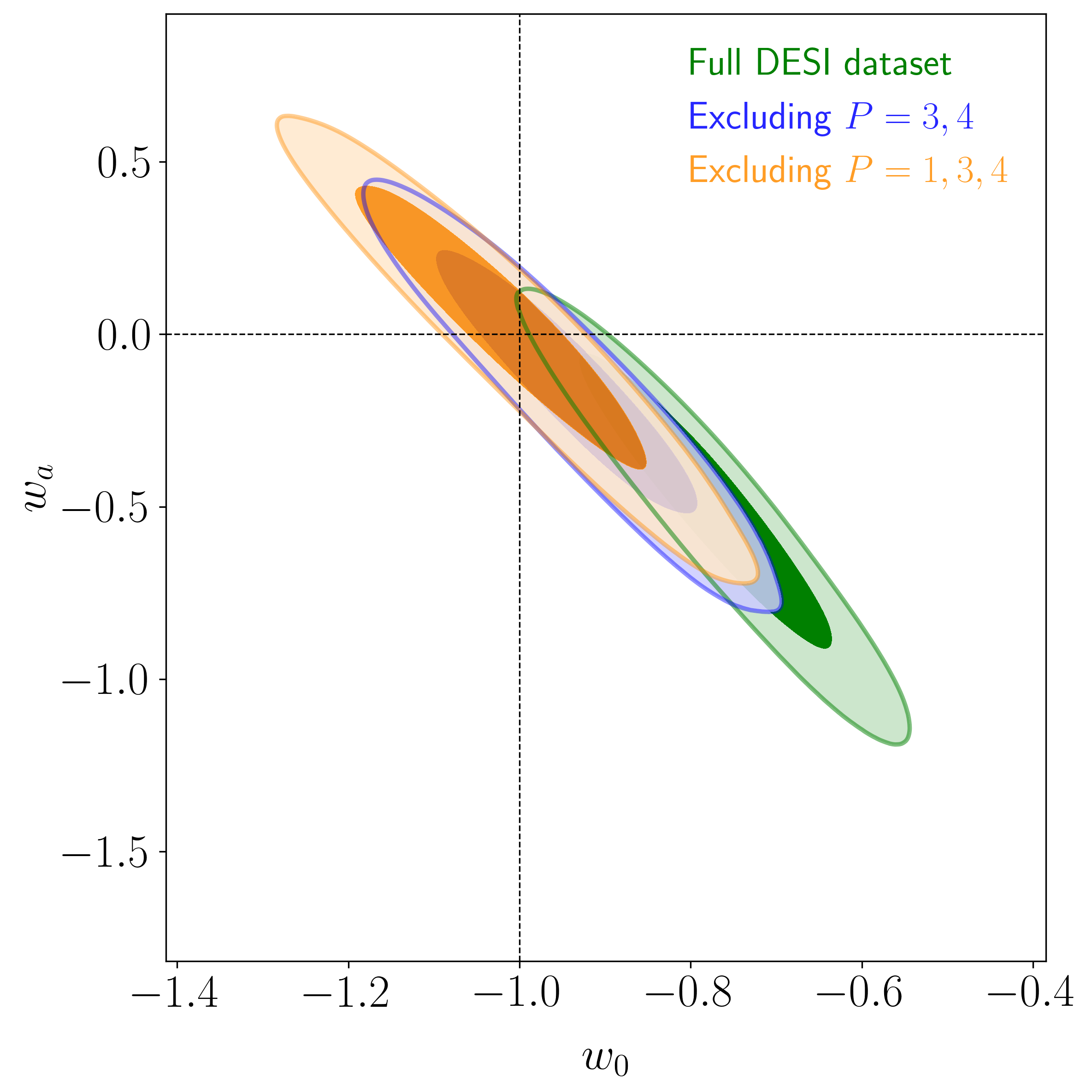}
\caption{Contour plots showing the $1\sigma$ and $2\sigma$ confidence regions for the equation of state parameters $w_0$ and $w_a$. Results are presented for the full dataset (green), the dataset excluding $P = 3$ and $P = 4$ (blue), and the dataset excluding $P = 1$, $P = 3$, and $P = 4$ (orange). }
\label{fig:contours-2sigma}
\end{figure}

In Fig.~\ref{fig:contours-2sigma}, we show the contour plots at $1$ and $2,\sigma$ for the full dataset (shown in green), as well as the contours obtained when $P = 3$ and $P = 4$ are excluded (in blue), and when $P = 1$, $P = 3$, and $P = 4$ are excluded (in orange). Figure~\ref{fig:contours-2sigma} clearly shows that excluding the identified outliers aligns the results with $\Lambda$CDM predictions when at least two specific points are removed. A similar result is found when all three outliers are excluded. The best-fit parameter values for these cases are summarized in Table~\ref{tab:best_fit_parameters}.

\textbf{\emph{Conclusions.}} In this work, we investigated the internal consistency of the DESI Year-1 BAO data by using the internal robustness algorithm of Amendola et al.~\cite{Amendola:2012wc} that  systematically analyzes subsets of the dataset in question in a fully Bayesian manner.  In particular, we identified the data points $P = 1,\,3,\,4$ as outliers. Moreover, our results show that while the full DESI dataset exhibits deviations from $\Lambda$CDM, excluding these specific data points brings the values of the equation of state parameters into agreement with the standard cosmological model. 

To further test our results, we repeated our analysis using an alternative dark energy equation of state, as proposed for instance by \cite{Giare:2024gpk}. In that work, the authors explored several parameterizations of the dark energy equation of state and found that the Barboza-Alcaniz model \cite{Barboza:2008rh} provides best-fit values of $w_0 = -0.848 \pm 0.054$ and $w_a = -0.38 {}^{+0.15}_{-0.13}$ when combining Planck, DESI, and PantheonPlus data. Applying this parameterization of $w(a)$ to our analysis, we again identified the same outliers. Specifically, when considering the full DESI dataset, we obtain $w_0 = -0.796 \pm 0.097$ and $w_a = -0.431 \pm 0.284$, whereas only excluding ${\bf P} = 3$ and $4$ yields $w_0 = -0.962 \pm 0.102$ and $w_a = -0.082 \pm 0.265$.

Our analysis demonstrates that the results are robust regarding the chosen DE model, highlighting the reliability of our approach.  It is fair to mention that outliers are data points that deviate significantly from the central trend (e.g., mean, median), warranting extra caution before drawing definitive conclusions. In this context, \cite{Mello:2024tor} introduced the non-Gaussian Surprise statistic as a tool to evaluate concordance between DESI and other cosmological datasets, finding notable tensions with Pantheon+ \cite{Scolnic:2021amr} and SH0ES \cite{Riess:2021jrx} data in both $\Lambda$CDM and $w_0w_a$CDM models.

However, whether the observed outliers imply systematic errors in the data or point towards new physics remains an open question that requires further investigation. 
Notably, DESI data have been extensively tested for potential systematics in prior studies, and these tests consistently suggest that systematic effects, both observational and theoretical, are under control \cite{DESI:2024fxb, Chen:2024tfp, DESI:2024tlb, Pinon:2024wzd, DESI:2024rkp, DESI:2024sho, DESI:2024ude, DESI:2024sbq}.
However, our analysis in this work seems to hint towards the existence of outliers in the data, thus still leaving open the possibility for new late time cosmological dynamics or other sources of systematics.

\noindent \textbf{\emph{Code availability:}} The code will be available upon publication.\footnote{\url{https://github.com/domenicosapone/}}

\noindent \textbf{\emph{Acknowledgements.}} DS acknowledges financial support from Fondecyt Regular N.~1210876. SN acknowledges support from the research project PID2021-123012NB-C43 and the Spanish Research Agency (Agencia Estatal de Investigaci\'on) through the Grant IFT Centro de Excelencia Severo Ochoa No CEX2020-001007-S, funded by MCIN/AEI/10.13039/501100011033. 

\bibliography{references}

\end{document}